# Massively-multiplexed generation of Bell-type entanglement using a quantum memory

Michał Lipka[1,2,4], Mateusz Mazelanik[1,2,4], Adam Leszczyński[1,2], Wojciech Wasilewski[1] & Michał Parniak[1,3 ✉]

High-rate generation of hybrid photon-matter entanglement remains a fundamental building block of quantum network architectures enabling protocols such as quantum secure communication or quantum distributed computing. While a tremendous effort has been made to overcome technological constraints limiting the efficiency and coherence times of current systems, an important complementary approach is to employ parallel and multiplexed architectures. Here we follow this approach experimentally demonstrating the generation of bipartite polarization-entangled photonic states across more than 500 modes, with a programmable delay for the second photon enabled by qubit storage in a wavevector-multiplexed cold-atomic quantum memory. We demonstrate Clauser, Horne, Shimony, Holt inequality violation by over 3 standard deviations, lasting for at least 45 $\mu$s storage time for half of the modes. The ability to shape hybrid entanglement between the polarization and wavevector degrees of freedom provides not only multiplexing capabilities but also brings prospects for novel protocols.

[1] Centre for Quantum Optical Technologies, Centre of New Technologies, University of Warsaw, Banacha 2c, Warsaw 02-097, Poland. [2] Faculty of Physics, University of Warsaw, Pasteura 5, Warsaw 02-093, Poland. [3] Niels Bohr Institute, University of Copenhagen, Blegdamsvej 17, Copenhagen 2100, Denmark. [4] These authors contributed equally: Michał Lipka, Mateusz Mazelanik. ✉email: m.parniak@cent.uw.edu.pl

Efficient and robust entanglement generation between distant parties remains one of the most fundamental steps towards practical implementations of quantum enhanced protocols[1–15], enabling amongst others quantum secure communication[1,10,16–20] or distributed quantum computing[21]. The optical domain remains a platform of choice for current state-of-the-art experimental demonstrations of quantum communication protocols; however, a fundamental exponential transmission loss limits the attainable distance between the parties to several tens of km[22,23]. While direct amplification of the quantum states is forbidden by the no-cloning theorem, noise-tolerant quantum repeaters have been proposed to distribute entanglement over extended distances by generating entanglement between intermediate stations (repeater nodes) and performing the entanglement swapping protocol[24,25]. Quantum memories play a significant role in such a task facilitating adaptive strategies, and enabling hierarchical architecture of the network with neighbor pairs of nodes waiting for each other, to accomplish entanglement the generation step before trying the entanglement swap.

While there are proposals for memory-less error-correction-based schemes[26,27], a significant technological advancement is still required, making the quantum-memory-based repeaters the most promising solution for near-term quantum networks. The performance of such repeaters heavily relies on the efficiency and lifetime of quantum memories, and a significant effort has been devoted to their improvement[28,29]. State-of-the-art quantum memories with either high retrieval efficiency[30–34] or long storage times[35–38] have been demonstrated. Even though, practical quantum repeater networks will most likely require a combination of those results with each other in a parallelized or multiplexed architecture[3,7,39–42], in particular facilitated by multimode quantum memories[42–44] and optical switching networks. Importantly, there has been a tremendous progress in developing either spatially[44,45], temporally[40,46–48], spectrally[49] or hybrid[42] multimode quantum memories. Recently demonstrated wavevector multiplexing constitutes a viable alternative offering both an unprecedented number of modes and versatile in-memory processing capabilities[43,50]. In our group, we have recently proposed and analyzed theoretically a scheme to use the wavevector multiplexing paradigm with a cold-atomic quantum memory as a feasible platform for a quantum repeater[51]. Roughly speaking, during entanglement generation the large number of modes allows for successful completion of this first step in almost all cases.

Here, we experimentally demonstrate a high-rate bipartite polarization entanglement generation employing ca. $M \approx 550$ pairs of photonic modes of the wavevector multiplexed quantum memory (WV-MUX-QM), with the possibility of delaying a second photon in pair for up to tens of μs. The entanglement is verified by a mode-resolved Bell state measurement certifying non-locality of generated states. The wavevector multiplexing approach is compatible with modern two-photon quantum repeater protocols robust to global phase fluctuations, and particularly suitable with promising multi-mode quantum optical channels employing multicore fibers[52,53] or free-space transmission[54]. Our scheme constitutes an important step towards quasi-deterministic entanglement generation[51], which for the achieved parameters would improve the quantum communication rate roughly $M$-fold.

We note that the Bell state generation has been demonstrated before in few-mode quantum memories[55,56], and recently with a similar mode conversion setup and a long lifetime of 1 ms[57], albeit only in three modes.

## Results

**Multimode generation of Bell states.** Robust quantum repeater protocols based on two-photon interference require high-rate generation of photon-atom entanglement[58–60] of the form $(S_H^\dagger a_H^\dagger + S_V^\dagger a_V^\dagger)/\sqrt{2}$, where $a_H^\dagger$ ($a_V^\dagger$) corresponds to a photonic creation operator for horizontal (vertical) polarization and $S_H^\dagger$ ($S_V^\dagger$) is the creation operator for an atomic excitation (spin-wave) associated with the emission of an $H$ (a $V$) photon. In a basic scenario, two atomic ensembles labeled $H$ and $V$ generate horizontally and vertically polarized photons, respectively, to further have them combined on a polarizing beamsplitter; however, such a scheme requires four atomic ensembles per repeater node (a two-ensemble interface per each neighbor) and complicates any multiplexing or multimode platform. Alternatively, with a light-matter interface supporting $2M$ single-polarization modes, $M$ photonic modes can have $H$ polarization ($V$ polarization) and be superimposed onto the other orthogonally polarized $M$ modes, provided the modes are coherent with each other. While we introduce this idea in the context of a wavevector multiplexed quantum memory (WV-MUX-QM), such an approach could be suited to other multimode systems.

The WV-MUX-QM is a high-optical-density cold-atomic-ensemble-based quantum memory with several hundreds of wavevector or equivalently photonic emission angle modes. The modes are interfaced via spontaneous off-resonant Raman scattering to probabilistically generate a two-mode squeezed state of atomic excitations (spin-waves), and a write-out (Stokes) photons in each mode, as detailed in the ref.[43]. The probability of generating a single pair of a spin-wave and a write-out photon per mode shall be denoted by $\chi \ll 1$. Importantly, the intermode coherence of WV-MUX-QM memory has been previously verified[61].

**Bell state generation across many modes.** Experimentally, we employ the wavevectors (or emission angles) of photons as the degree of freedom for the $M$ modes. The memory is interfaced with spatially large write/read beams with a well defined and the fixed angle corresponding to wavevectors $\mathbf{k_W}$ and $\frac{\mathbf{k_R}}{|\mathbf{k_R}|} = -\frac{\mathbf{k_W}}{|\mathbf{k_W}|}$ for write and read beam, respectively. In the memory writing process, a spin-wave–photon pairs across many angular modes are generated. The quantum state of a single pair of this kind can be described as a coherent superposition of plane-wave contributions:

$$|\psi\rangle = \int_A d\mathbf{k_w} S_V^\dagger(\mathbf{k_W} - \mathbf{k_w}) a_V^\dagger(\mathbf{k_w})|vac\rangle, \quad (1)$$

where $\mathbf{k_W} - \mathbf{k_w} = \mathbf{K}$ represents a unique spin-wave wavevector for each (further called write-out—$w$) generated photon's wavevector $\mathbf{k_w}$ and $A$ represents the rectangular field of view in the wavevector space. Additionally, the write-out photons coming from the memory have circular polarization that is further transformed to vertical one, which we explicitly denote by $V$ subscript.

In a single write laser shot many pairs can be generated, and the full quantum state of the memory and write-out field is:

$$|\Psi\rangle = |vac\rangle + \sqrt{\chi_M}|\psi\rangle + \mathcal{O}(\chi_M), \quad (2)$$

where $\chi_M$ is a total (across $M$ modes) pair generation probability, which for small $\chi$ is just $\chi_M \approx \chi M$. The state $|\Psi\rangle$ describes many spin-wave–photon pairs distributed across all possible modes including multiphoton contributions to a single mode. However, as the probability of such an event is $\chi^2$ (for a chosen mode), for small $\chi$ they are not likely to happen, even with relatively high average of total excitations per shot. Therefore, we can focus on a single photon–spin-wave state $|\psi\rangle$ from which we generate the WV-MUX Bell state.

To generate the polarization degree of freedom (DoF) Bell state, we split the field of view containing $2M$ modes in half and superimpose the two resulting regions. Prior to the superimposition, one half is sent through half-waveplate to change polarization of the write-out photons form vertical ($V$) to horizontal ($H$). After the superimposition we end up with a WV-MUX polarization DoF Bell-like state:

$$|\psi_B\rangle = \int_{A/2} d\mathbf{k_w} \left( S_H^\dagger(\mathbf{K}) a_H^\dagger(\mathbf{k_w}) + e^{i\varphi_w(\mathbf{k_w})} S_V^\dagger(\mathbf{K}) a_V^\dagger(\mathbf{k_w}) \right) |vac\rangle, \quad (3)$$

where the optical phase between generated write-out $H$ and $V$ photons with $\mathbf{k_w}$ wavevector is $\varphi_w(\mathbf{k_w})$. The spin waves (atomic excitations) can be read-out after a programmable delay:

$$S_X^\dagger(\mathbf{K}) \to \int d\mathbf{k_r} \phi(\mathbf{k_r}, \mathbf{K} + \mathbf{k_R}) b_X^\dagger(\mathbf{k_r}), \quad (4)$$

where $X \in \{H, V\}$ and $b_X^\dagger(\mathbf{k_r})$ denotes a creation operator for the read-out photon with polarization $X$ and wavevector $\mathbf{k_r}$, and $\phi(\mathbf{k_r}, \mathbf{k_w})$ characterizes the correlation between write-out and read-out photons. The operation brings our state to:

$$|\psi_B\rangle = \int_{A/2} d\mathbf{k_w} d\mathbf{k_r} \phi(\mathbf{k_r}, \mathbf{k_w}) \mathfrak{B}^\dagger(\mathbf{k_r}, \mathbf{k_w}) |vac\rangle, \quad (5)$$

$$\mathfrak{B}^\dagger(\mathbf{k_r}, \mathbf{k_w}) = b_H^\dagger(\mathbf{k_r}) a_H^\dagger(\mathbf{k_w}) + e^{i\varphi(\mathbf{k_r}, \mathbf{k_w})} b_V^\dagger(\mathbf{k_r}) a_V^\dagger(\mathbf{k_w}), \quad (6)$$

where we introduced a WV-MUX Bell state creation operator $\mathfrak{B}^\dagger(\mathbf{k_r}, \mathbf{k_w})$, with $\varphi(\mathbf{k_r}, \mathbf{k_w}) = \varphi_w(\mathbf{k_w}) + \varphi_r(\mathbf{k_r})$ and $\varphi_r(\mathbf{k_r})$ being an additional phase difference for orthogonally polarized read-out photons.

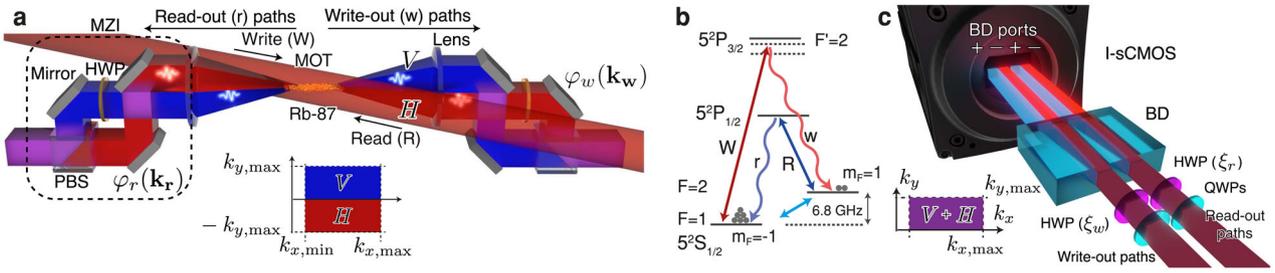

**Fig. 1 Wavevector-multiplexed quantum memory as a Bell-state generator. a** Photonic modes of write-out photons emitted from the atomic cloud (MOT) are divided by their emission angle into horizontally ($H$, red beam) and vertically ($V$, blue beam) polarized parts which are further superimposed creating polarization Bell states across many modes. Such an operation of the memory is facilitated with two Mach–Zehnder interferometers (MZI) placed in the write-out and read-out paths of the atomic emission. HWP half-wave plate set for $H \leftrightarrow V$ conversion, PBS polarizing beamsplitter. **b** Double lambda configuration of Rb-87 energy levels constituting the light-matter interface of WV-MUX-QM. W write beam, R read beam, w write-out photons, r read-out photons. **c** Multimode Bell-state measurement (BSM) between the write-out and read-out photons generated in the WV-MUX-QM. I-sCMOS intensified scientific CMOS camera facilitating single-photon measurements, BD beam displacer, HWP half-wave plate, QWP quarter-wave plate.

We note that with anti-collinear write and read beam configuration ($\frac{\mathbf{k_R}}{|\mathbf{k_R}|} = -\frac{\mathbf{k_W}}{|\mathbf{k_W}|}$) the write-out–read-out wavevectors on average satisfy $\frac{\mathbf{k_w}}{|\mathbf{k_w}|} = -\frac{\mathbf{k_r}}{|\mathbf{k_r}|}$. Since the read-out of a spin-wave is facilitated by light-atom interaction in a finite ensemble, the angular spread of read-out photons around specific $\mathbf{k_r}$ corresponding to a registered $\mathbf{k_w}$ is inversely proportional to the transverse size of the atomic cloud. Assuming Gaussian profile for the transverse atomic density, we get a Gaussian mode function (conditional on the choice of the write-out wavevector $\mathbf{k_w}$) for the read-out photon with $\mathbf{k_r}$ wavevector, described by the correlation function:

$$\phi(\mathbf{k_r}, \mathbf{k_w}) = \mathcal{N} \exp\left[ (-\mathbf{k_w} - \mathbf{k_r})^T \frac{\mathcal{C}^{-1}}{2} (-\mathbf{k_w} - \mathbf{k_r}) \right],$$
$$\mathcal{N} = \sqrt{\frac{1}{2\pi\sqrt{\det \mathcal{C}}}}, \quad (7)$$

with the covariance matrix $\mathcal{C}$, that determines the correlation strength between write-out and read-out photons in the $\mathbf{k}$-space. The correlation size gives the number of memory modes in terms of Schmidt decomposition:

$$M \approx \frac{\alpha}{4\sqrt{\lambda_1 \lambda_2}} A,$$

where $\alpha$ is a geometric factor, which for a rectangular area with transversally-Gaussian correlation is $\alpha = 0.565$[43], and $A = L_x \times L_y$ denotes the observed rectangular area in the wavevector space (note that in total we observe $2A$ which is divided into $H$ and $V$ part) and $\lambda_1$, $\lambda_2$ correspond to eigenvalues of the $\mathcal{C}$ matrix (e.g., $\lambda = \sigma^2$ for one dimensional Gaussian parameter $\sigma$).

Finally, we can rewrite $|\psi_B\rangle$ in terms of the two-photon (write-out–read-out pair) subspace and separate the wavevector and polarization photonic DoFs, obtaining:

$$|\psi_B\rangle = \int_{A/2} d\mathbf{k_w} d\mathbf{k_r} \phi(\mathbf{k_r}, \mathbf{k_w}) |\mathbf{k_r}\rangle |\mathbf{k_w}\rangle |\Phi(\varphi(\mathbf{k_r}, \mathbf{k_w}))\rangle \quad (8)$$

with a polarization DoF state:

$$|\Phi(\varphi)\rangle = \frac{1}{\sqrt{2}} \left[ |H\rangle_r |H\rangle_w + \exp(i\varphi) |V\rangle_r |V\rangle_w \right]. \quad (9)$$

We note that the state given by Eq. (8) has a non-trivial interdependence between the wavevector and polarization DoF via the wavevector-dependent phase $\varphi_w(\mathbf{k_w}) + \varphi_r(\mathbf{k_r})$ (see "Methods" section), which is experimentally feasible to be arbitrarily shaped e.g., by placing a spatial light modulator (SLM) in the far-field of the atomic cloud.

Throughout this work we choose the coordinates so that the photons are correlated as follows:

$$\phi(\mathbf{k_r}, \mathbf{k_w}) \equiv \phi(x_w - x_r, y_w - y_r)$$
$$\propto \exp[-(y_w - y_r)^2/(2\sigma_y^2)] \times \exp[-(x_w - x_r)^2/(2\sigma_x^2)], \quad (10)$$

where we denote $\mathbf{k_i} = (x_i, y_i)$; $i \in \{r, w\}$.

*Experimental Bell-state preparation.* To experimentally generate a state given by Eq. (8) across $M \approx 550$ modes, we employ a WV-MUX-QM with a specifically designed Mach–Zehnder interferometers (MZI) placed in the far-field of the atomic cloud (see "Methods" section). A simplified experimental setup is depicted in Fig. 1. Each MZI allows dividing the emission cone (wavevector range) of write-out (read-out) photons to $H$ and $V$ modes and to superimpose the two parts. Furthermore, the wavevector-dependent phase $\varphi_w(\mathbf{k_w})$, $\varphi_r(\mathbf{k_r})$ can be shaped by tilting one of the MZI mirrors. The MZI consists of three mirrors, a half-wave plate (HWP) and a polarizing beam-splitter (PBS). All modes have initially $H$ polarization which is not reflected at PBS. The first mirror reflects half of the emission cone (half of the wavevector modes), which is then routed by the second mirror into a port of PBS and transmitted through. The other half of the emission cone passes through HWP, which rotates the polarization to $V$, and is reflected by the third mirror into the second port of the PBS. $V$-polarized half is reflected at the PBS, hence the two parts are superimposed in wavevector (spatial) DoF. Importantly, both parts undergo the same number of reflections. Effectively, for each pair of modes MZI connects two wavevector modes into a product of a single wavevector mode and two polarization modes.

Each MZI works as a mode converter between the wavevector and polarization DoF. The wavevector DoF enables parallelization or multiplexing for efficient quantum repeater protocols[51], while the polarization DoF facilitates Bell state generation and enables robust two-photon protocols for entanglement creation and connection between repeater nodes[58–60].

We note that the conversion between wavevector (or spatial) and polarization DoFs has been demonstrated before, also in the context of Bell state generation[41,48,57,62], albeit with a few-mode system.

**Bell-state measurement.** To quantify the entanglement of experimentally generated states we perform a wavevector-resolved Bell-state measurement (BSM) with the write-out and read-out arm corresponding to the two parties—Alice and Bob—usually considered in the Bell setting. With a linear MZI phase, we select measurement bases for Alice and Bob lying on the equator of the Bloch sphere. Such a choice ensures the BSM visibility remains constant regardless of the wavevector-dependent phase. The local Alice and Bob generalized measurement operators $\{\Theta(\mathbf{k})\}_\mathbf{k}(\xi)$ are given by

$$\Theta(\mathbf{k}, \xi) = |\mathbf{k}\rangle\langle\mathbf{k}| \otimes \Pi_\xi, \quad (11)$$

with $\Pi_\xi = \hat{\sigma}_x \cos\xi + \hat{\sigma}_y \sin\xi$, where $\hat{\sigma}_x, \hat{\sigma}_y$ are the Pauli operators and $\xi$ parametrizes the measurement (e.g., $\xi = 0, \xi = \pi/2$ correspond to a measurement in diagonal and circular polarization bases, respectively).

Write-out and read-out paths undergo a projective polarization measurement on a beam-displacer (BD) with $\xi$ adjusted by half-wave plates (HWP). The two output ports ± of the BD are observed with a single-photon sensitive I-sCMOS camera[63], which resolve individual wavevector modes with 1 px corresponding to 2.38 rad/mm (see "Methods" section). Importantly, some of the events registered with the camera correspond to dark counts, photons from the multiexcitation component $\mathcal{O}(\chi)$ or misalignment of the experimental setup. To model those imperfections we assume a depolarizing channel[51] over the polarization DoF, which transforms:

$$|\Phi(\varphi)\rangle\langle\Phi(\varphi)| \to \frac{1-\mathcal{V}}{4}\hat{\mathbb{I}} + \mathcal{V}|\Phi(\varphi)\rangle\langle\Phi(\varphi)|, \quad (12)$$

where the visibility $\mathcal{V}$ in general depends on the wavevectors of write-out and read-out photons.

*Expected value and Bell parameter.* In the experimental demonstration, we select a linear MZI phase

$$\varphi_r(\mathbf{k_r}) = \mathbf{a_r} \cdot \mathbf{k_r} + \varphi_0, \quad (13)$$

$$\varphi_w(\mathbf{k_w}) = \mathbf{a_w} \cdot \mathbf{k_w}, \quad (14)$$

which for maximally correlated write-out and read-out wavevectors $\mathbf{k_w} = \mathbf{k_r} = \mathbf{k}$ allows us to observe a family of Bell-like states $|\Phi(\mathbf{a} \cdot \mathbf{k} + \varphi_0)\rangle$ with $\mathbf{a} = \mathbf{a_w} + \mathbf{a_r}$.

Let us calculate the average outcome of local Alice's and Bob's measurements for a single pair of write-out and read-out wavevectors $\Theta(\mathbf{k_w}, \xi_w) \otimes \Theta(\mathbf{k_r}, \xi_r)$. Post-selecting outcomes with a registered pair of photons at $\mathbf{k_r}$ and $\mathbf{k_w}$, we get the expected value to be

$$E(\mathbf{k_r}, \mathbf{k_w}; \xi_r, \xi_w) = \mathcal{V}(\mathbf{k_r}, \mathbf{k_w})\cos(\varphi(\mathbf{k_r}, \mathbf{k_w}) + \xi_r + \xi_w), \quad (15)$$

where the cosine is given by the trace over polarization DoF: $\text{Tr}\left[(\Pi_{\xi_w} \otimes \Pi_{\xi_r})|\Phi(\varphi(\mathbf{k_r}, \mathbf{k_w}))\rangle\langle\Phi(\varphi(\mathbf{k_r}, \mathbf{k_w}))|\right]$. Selecting two bases $\mathcal{A} = \{\xi_w^{(1)}, \xi_w^{(2)}\}$, $\mathcal{B} = \{\xi_r^{(1)}, \xi_r^{(2)}\}$ for Alice and Bob each, respectively, we can formulate the Bell parameter in the wavevector space:

$$\mathcal{S}(\mathcal{A}, \mathcal{B}, \mathbf{k_r}, \mathbf{k_w}) = \sum_{i,j=1}^{2}(1 - 2\delta_{i,j}\delta_{i,2})E(\mathbf{k_r}, \mathbf{k_w}; \xi_r^{(i)}, \xi_w^{(j)}). \quad (16)$$

For an optimal selection of bases

$$\mathcal{A}^*(\mathbf{k_r}, \mathbf{k_w}), \ \mathcal{B}^*(\mathbf{k_r}, \mathbf{k_w}) = \arg\max_{\mathcal{A},\mathcal{B}}|\mathcal{S}(\mathcal{A}, \mathcal{B}, \mathbf{k_r}, \mathbf{k_w})|, \quad (17)$$

the Bell parameter

$$\mathcal{S}^*(\mathbf{k_r}, \mathbf{k_w}) = \max_{\mathcal{A},\mathcal{B}}|\mathcal{S}(\mathcal{A}, \mathcal{B}, \mathbf{k_r}, \mathbf{k_w})| \quad (18)$$

depends only on the visibility

$$\mathcal{S}^*(\mathbf{k_r}, \mathbf{k_w}) = 2\sqrt{2}\mathcal{V}(\mathbf{k_r}, \mathbf{k_w}). \quad (19)$$

To violate Clauser, Horne, Shimony, Holt (CHSH) inequality ? and certify non-classical correlations one needs $\mathcal{V}(\mathbf{k_r}, \mathbf{k_w}) > 1/\sqrt{2}$.

*Visibility.* BSM visibility indirectly quantifies the quality of the generated entangled state for further entanglement distillation protocols. Entanglement of distillation, i.e., the number of maximally entangled states that can be distilled per a copy of the generated state, would provide a resource-oriented characterization; however, it is difficult to calculate in a generic case. Optimistically, one may use another entanglement monotone such as entanglement of formation, concurrence or negativity to upper bound the entanglement of distillation. For a Werner state, given by RHS of Eq. (12), all those monotones can be calculated analytically[64] and depend only on the visibility. Furthermore, concurrence and negativity are in this case linear functions of $\mathcal{V}$, hence we will further focus on the visibility as a figure of merit quantifying the entanglement of generated states.

In addition to experimental setup imperfections, the influence of multiphoton excitations and noise either from dark counts or stray photons constitutes a fundamental factor limiting the visibility. Importantly, these factors also affect the Glauber second-order intensity cross-correlation between write-out and read-out photons $g^{(2)}(\mathbf{k_r}, \mathbf{k_w})$. As shown in "Methods" section, the visibility can be written as:

$$\mathcal{V}(\mathbf{k_r}, \mathbf{k_w}) = \frac{g^{(2)}(\mathbf{k_r}, \mathbf{k_w}) - 1}{g^{(2)}(\mathbf{k_r}, \mathbf{k_w}) + 1}. \quad (20)$$

Interestingly, this result facilitates estimating BSM visibility via $g^{(2)}$ function measurements which do not require the BSM setup. Importantly, with the BSM setup in-place, the $g^{(2)}$ function lets us observe wavevector-resolved interference in the correlation patterns, facilitating observation of the phase profile at MZIs (see the Supplementary Material Sec. S1 and S2 for more details about the wavevector space correlations).

**Experimental BSM.** To observe the mode-resolved Bell parameter $\mathcal{S}(\mathcal{A}, \mathcal{B}, \mathbf{k_r}, \mathbf{k_w})$ we shall look at maximally correlated pairs of write-out and read-out wavevectors $\arg\max|\phi(\mathbf{k_w}, \mathbf{k_r})|^2$. As detailed in the Supplementary Material Sec. S3, for each point in sum coordinates $[y_s \equiv (y_w + y_r)/2, x_s \equiv (x_w + x_r)/2]$, we bin the coincidences in a rectangular region $2n\sigma_x \times 2n\sigma_y$ around $(0, 0)$ point in difference $(y_w - y_r, x_w - x_r)$ coordinates. Importantly, this affects the second-order correlation function and thus also the BSM visibility which reads:

$$\mathcal{V} = \left[1 + \frac{2\chi(1+\tilde{B}_w/\chi)\{1+\tilde{B}_r/[\chi\eta_r]\}}{F(n)}\right]^{-1}, \quad (21)$$

where the binning factor is

$$F(n) = \mathrm{Erf}\,(n/\sqrt{2})^2/(\alpha n^2) \quad (22)$$

and where $\tilde{B}_w = B_w/M$ ($\tilde{B}_r = B_r/M$) denotes the noise probability per mode in the write-out (read-out) arm, before detection. The visibility is expected to be uniform in sum coordinates $(x_s, y_s)$. For further analysis we choose $n = 1$ ($F(n) \approx 0.825$) as a trade-off between decreasing the visibility (as $n$ grows) and gathering a larger statistics for each point. The choice of Alice and Bob bases $\mathcal{A}, \mathcal{B}$ is optimal for a subset of wavevector modes. While, we note that a constant MZI phase would enable a choice of bases simultaneously optimal for all modes, linear phase provides experimentally feasible characterization in terms of the BSM visibility. Figure 2 depicts subsequent analysis stages leading to the Bell parameters in sum coordinates $\mathcal{S}(x_s, y_s)$. For Fig. 2a–d we choose a single pair of bases $\xi_w^{(2)}, \xi_r^{(2)}$ and illustrate coincidence maps with different combinations of polarization measurement outcomes in write-out/read-out arms. Figure 2e–h depict expected values for the BSM measurement with each combination of Alice and Bob bases e.g., Fig. 2e corresponds to the following operation on maps depicted in Fig. 2a–d: $e = (a - b - c + d)/(a + b + c + d)$. Wavevector-resolved Bell parameter, depicted in Fig. 2i, is obtained according to Eq. (16) by taking a combination of BSM expected values. The linear MZI phase is clearly visible in all maps, modulating the results along $y_s$. Hence, we take an average of the Bell parameter along $x_s$. As depicted in Fig. 2j, the averaged Bell parameter $\langle \mathcal{S}(x_s, y_s) \rangle_{x_s}$ is sinusoidal with a fitted amplitude of 2.60 ± 0.19 yielding CHSH violation by over three standard deviations.

### BSM with memory

*BSM visibility.* The Bell parameter under a choice of optimal bases is directly proportional to the BSM visibility, making the visibility a viable figure of merit for the generated write-out–read-out bipartite states. Furthermore, from the perspective of entanglement distillation protocols, the BSM visibility determines the entanglement monotones measuring the ebit content of a single generated state. Importantly, entanglement distribution protocols such as quantum repeater nets can greatly benefit from the delayed release of the read-out photon, which allows improving the protocol success rate via, among others, hierarchical architectures[25,59] or multiplexing of the read-out photonic mode[51]. Hence, we perform a series of measurements with increasing memory time $t \in [0.3, 60.3]$ μs and with a very quickly changing linear MZI phase, which allows us to retrieve wavevector-resolved BSM visibility from the sinusoidal patterns in coincidence maps.

*Spin-wave decoherence.* Fundamentally, the BSM visibility at larger memory times is limited by the spin-wave decoherence. For WV-MUX-QM's light-matter interface, we select the so-called clock transitions insensitive to the first order to stray magnetic fields and fix the quantization axis by introducing external constant magnetic field along the cloud (z-axis); hence, the decoherence mechanism in our case mainly due to the thermal atomic motion. Inevitably, random displacement of atoms distorts the spatial structure of a spin-wave[43]. The decoherence quantified as an average overlap with the initial spin-wave state

$$|\langle S_\mathbf{k} | S_\mathbf{k}(t) \rangle|^2 \propto \exp(-t^2/\tau(\mathbf{k})^2) \quad (23)$$

is Gaussian with the characteristic time $\tau$ depending on the spin-wave wavevector and given by:

$$\tau(\mathbf{k}) = \gamma/|\mathbf{k}|, \quad (24)$$

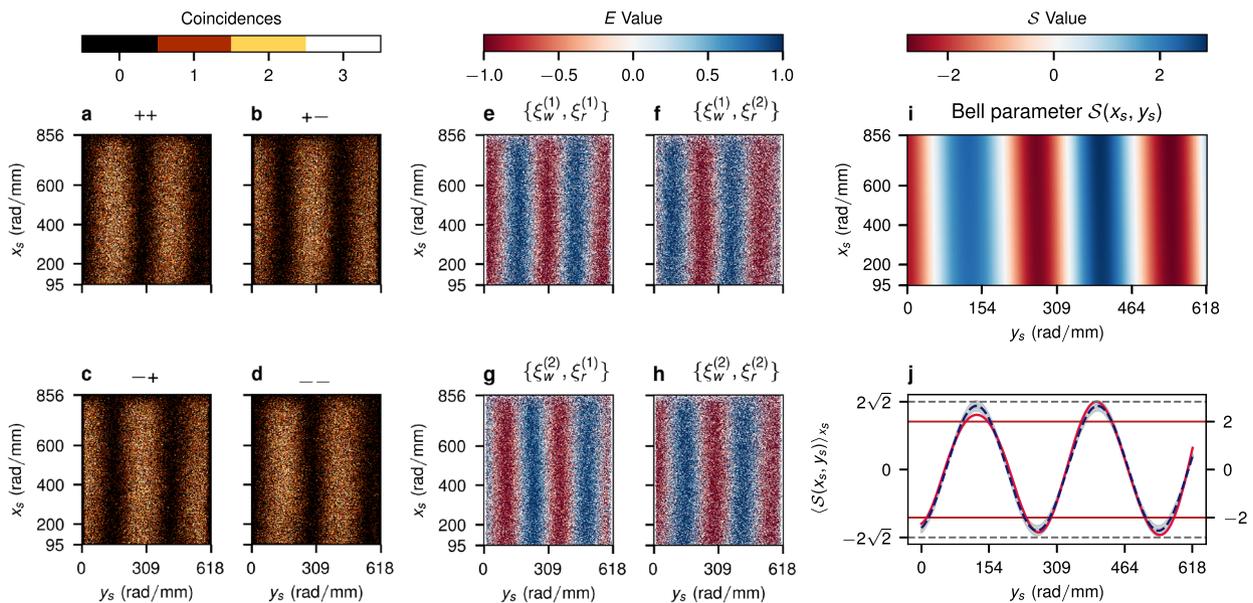

**Fig. 2 Wavevector-resolved Bell-state measurement between write-out and read-out photons. a–d** Coincidences between $(s_r, s_w) = \{(+, +), (+, -), (-, +), (-, -)\}$ ports, respectively, with a single pair of Alice and Bob bases $\{\xi_w^{(2)}, \xi_r^{(2)}\}$. **e–h** Expected values of a measurement for subsequent combinations of $\{\xi_w^{(1)}, \xi_w^{(2)}\}$ and $\{\xi_r^{(1)}, \xi_r^{(2)}\}$ bases. Each expected value is a linear combination of coincidences between ± ports. **i** Bell parameter $\mathcal{S}(x_s, y_s)$ retrieved by fitting a sine function (with $x_s$ and $y_s$ spatial frequencies) to the expected value for each combination of bases and taking the Bell linear combination of the results. **j** Bell parameter averaged over $x_s$ (red curve) along fitted sine (blue dashed curve) with amplitude 2.60 ± 0.19 indicating CHSH violation by more than three standard deviations $\sigma$. Subsequent shadings correspond to $\sigma$ and $2\sigma$. Solid lines indicate CHSH inequality violation threshold.

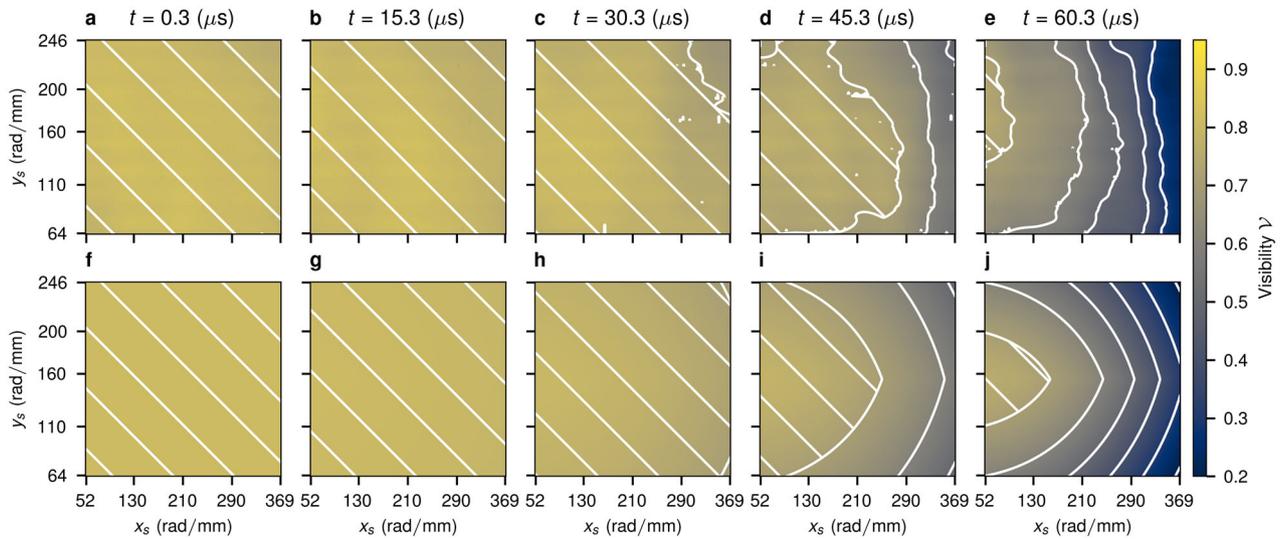

**Fig. 3 Evolution of visibility among many modes.** Figure presents Bell state visibility $\mathcal{V}(x_s, y_s; t)$ retrieved from coincidences patterns measured with a quickly changing linear MZI phase $\varphi(\mathbf{k}) \propto y_r \propto y_s$. Presented data covers a selected range of observed wavevectors and subsequently larger memory storage times $t$. **a–e** Experimental data. **f–j** Visibility model best fit. Hatched regions correspond to visibility $\mathcal{V} > 1/\sqrt{2} \approx 0.707$ yielding CHSH violation. Experimental data and model fits are averaged over four combinations of measurement ports $(s_r, s_w) = \{(+, +), (+, -), (-, +), (-, -)\}$. Before model fitting, experimental data is Gaussian filtered to reduce noise. Some of the raw data is also presented in Supplementary Material Sec. S5.

$$\gamma = \sqrt{m/(k_B T)}, \quad (25)$$

with $\gamma$ depending on the atomic mass of $^{87}$Rb $m$, cloud temperature $T$ and Boltzmann constant $k_B$. Spin-wave decoherence can be accounted for by plugging into Eq. (21) a time-dependent retrieval efficiency

$$\eta_r(t, \mathbf{k}) = \eta_r(0) \exp(-t^2/\tau(\mathbf{k})^2). \quad (26)$$

*Visibility model.* In our experimental setup the MZI used to divide the atomic emission cone into $H$ and $V$ polarized parts superimposes a mode with the lowest modulus wavevector $\min |\mathbf{k_H}|$ from the $H$ part onto the highest modulus wavevector $\max |\mathbf{k_V}|$ from the $V$ part and vice versa. As a consequence, spin-waves corresponding to different polarization parts decohere with a different rate $\tau(|\mathbf{k_H}|) \neq \tau(|\mathbf{k_V}|)$ effectively further deteriorating BSM visibility. Importantly, it could be amended by an improved MZI setup.

Let us denote $k_{\min} = \min(|\mathbf{k_H}|, |\mathbf{k_V}|)$, $k_{\max} = \max(|\mathbf{k_H}|, |\mathbf{k_V}|)$. Since longer wavevectors are associated with faster decoherence, let as also denote $\tau_{\min} = \tau(k_{\max})$, $\tau_{\max} = \tau(k_{\min})$ and $\Delta = \tau_{\min}^{-1} + \tau_{\max}^{-1}$. As demonstrated in "Methods" section, the additional visibility reduction is:

$$\tilde{\mathcal{V}}(t) = 2c_H(t)c_V(t)/(c_H^2(t) + c_V^2(t)) \\ = 1/\cosh(\Delta t^2/2). \quad (27)$$

With the increasing storage time the decoherence deteriorates the visibility. As detailed in Methods, the storage-time–dependent version of Eq. (21) can be written as:

$$\mathcal{V}(t) \approx \frac{\mathcal{V}_0}{1 + \frac{W}{\mathcal{V}_0}\exp[t^2/\tau(\mathbf{k})^2]}. \quad (28)$$

The total visibility is given by the product of Eq. (28) and Eq. (27).

*BSM visibility maps.* Experimentally, the BSM visibility can be retrieved from coincidence measurements with a quickly changing linear Mach-Zehnder inteferometer phase. The coincidence map for each combinations of measurement ports $(s_r, s_w) = \{(+, +),$ $(+, -), (-, +), (-, -)\}$ is smoothed with one-dimensional Gaussian filters ($\sigma = 1$ px for $y_i$ direction and $\sigma = 10$ px for $x_s$ direction) and for each $x_s$ a row of data along $y_s$ is selected. Each row is divided into overlapping 50 px segments with an 1 px step and to each segment we fit a model given by $a\cos(2\pi f_y y_s) + b$ and obtain visibility as $a/b$, which we assign to the $y_s$ coordinate of the visibility map given by the center position of the 50 px segment. Figure 3 depicts experimental data along a fitted visibility model given by the product of Eq. (28) and Eq. (27), assuming $\tau(\mathbf{k}) = \gamma/|\mathbf{k}|$. Hatched regions correspond to visibility above $1/\sqrt{2}$, which predicts CHSH violation while additional isolines are drawn at levels from 0.6 downward with a step of 0.1. The model fit yielded

$$\mathcal{V}_0 = 0.92 \pm 0.02, \tag{29}$$

$$W = 0.13 \pm 0.02, \tag{30}$$

$$\gamma = (6.26 \pm 0.29) \times 10^3 \ \mu\text{s rad mm}^{-1}, \tag{31}$$

with $\gamma$ corresponding to a temperature of $T = 47^{+5}_{-4} \ \mu\text{K}$. For completeness, we note that an independent measurement, described in Supplementary Material Sec. S4, yielded a consistent temperature of $T = 48^{+6}_{-5} \ \mu\text{K}$, the $\chi$-dependent noise $\tilde{B}r^{(\chi)} = 0.131 \pm 0.015$ and a read-out efficiency of $\eta_r(0) = 0.405 \pm 0.015$.

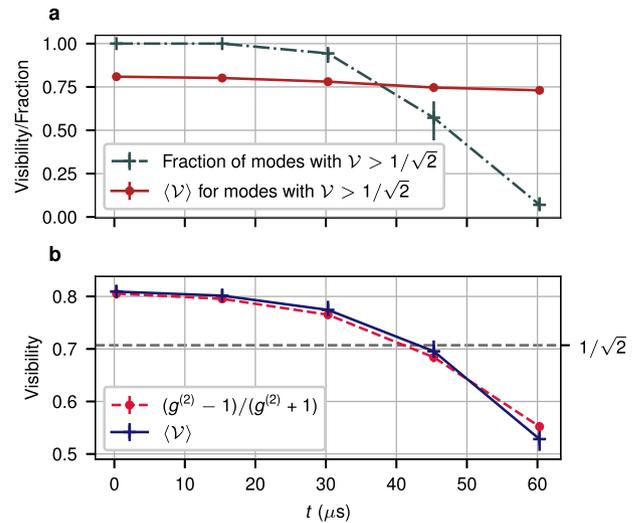

**Fig. 4 Persistence of high mean visibility.** Visibility has been averaged over wavevector modes $\langle \mathcal{V}(x_s, y_s; t) \rangle_{x_s, y_s}$ for increasing memory times $t$. **a** Blue curve represents the fraction of modes violating CHSH inequality, while the red curve corresponds to the visibility averaged over this subset of modes. **b** Visibility averaged over all modes (purple curve) yields good agreement with the model based on Glauber cross-correlation $g^{(2)}$ (red dashed curve, cf. Eq. (20)). Lines between points serve as guide to the eye. Errorbars represent extreme cases in the ensemble of wavevector-modes, with top/bottom of errobar corresponding to the best/worst visibility ± 1 s. d. obtained from the fit.

*Mode-averaged visibility.* Obtained visibility maps let us study the properties of an average memory mode. As depicted in Fig. 4a, up to ca. 30 μs nearly all modes would violate CHSH, while for 45 μs it remains true for half of the modes. The average visibility amongst the CHSH violating modes remains fairly constant with increasing memory time. Figure 4b depicts very good agreement of the visibility averaged over all modes $\langle \mathcal{V} \rangle$ with the mode-averaged prediction from the measured Glauber cross-correlation $\langle (g^{(2)}(t) - 1)/(g^{(2)}(t) + 1) \rangle$. Visibly, above 45 μs the performance of a significant number of modes has severely deteriorated. Nevertheless, modes with a lower modulus wavevector can have significantly longer lifetimes. Unfortunately, the number of modes with a given modulus wavevector $|\mathbf{k}|$ is roughly proportional to $|\mathbf{k}|$, hence there are few modes with a long lifetime.

*Wavevector-dependent decoherence rate.* For quantum memories utilizing a dense cold ensemble of atoms, it has been demonstrated that thermal motion constitutes the main source of spin-wave decoherence[38], provided the so-called clock transitions—in the first order robust to external magnetic field fluctuations—are utilized for light-matter interface. Characteristically, decoherence due to thermal motion leads to Gaussian temporal profile of the retrieval efficiency, as given by Eq. (26), with a characteristic time of $\tau(|\mathbf{k}|) = \gamma/|\mathbf{k}|$. A vast range of wavevector modes simultaneously utilized in the WV-MUX-QM provides a unique opportunity to precisely observe experimentally the decoherence time $\tau(|\mathbf{k}|)$ dependence on the wavevector length $|\mathbf{k}|$. As depicted in Fig. 5, experimental results are in a good agreement with the predictions of thermal-motion-induced decoherence.

## Discussion

We experimentally demonstrated the generation of polarization-entangled bipartite Bell states in ca. 550 photonic modes, with an inherent programmable delay for the second photon in a pair. Our approach harnesses hybrid atom-photon entanglement generation in a single cold high-density ensemble of Rb-87 atoms, which is the central part of the wavevector-multiplexed quantum memory (WV-MUX-QM). Ca. 1100 wavevector (angular emission) modes of the WV-MUX-QM are divided into $H$ and $V$ polarized photonic modes, which are superimposed pairwise forming 550 modes in a polarization superposition. Importantly, a write-out photon in one of those combined modes may have been created in an $H$ or a $V$ part, in each case having a different initial wavevector and thus being entangled with a collective atomic excitation—a spin-wave—in a different memory mode. The wavevectors of write-out and corresponding read-out photons are correlated with each other; hence, a write-out photon created in the $H(V)$ part of write-out emission cone is accompanied by a read-out photon also in the $H(V)$ part, albeit of the read-out emission cone. Hence, the read-out modes need to be super-imposed in the same way as write-out modes. This way, a write-out photon and a correlated read-out photon may be either both $H$ polarized or both $V$ polarized, constituting a polarization entangled state. Fundamentally, this concept realizes a conversion between wavevector and polarization degrees of freedom.

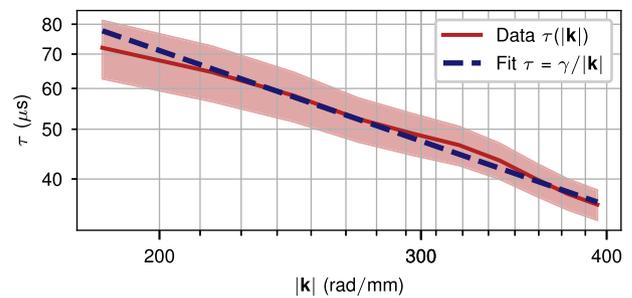

**Fig. 5 Dependence of the decoherence rate $\tau(|\mathbf{k}|)$ on the spin-wave wavevector length $|\mathbf{k}|$.** Data was obtained by selecting wavevectors with the same decoherence rate for both polarizations $c_H(t) = c_V(t)$ (i.e., $y_s \approx y_{\max}/2$) from the BSM visibility maps $\mathcal{V}(x_s, y_s; t)$. Data was binned by $|\mathbf{k}| = \sqrt{x_s^2 + y_s^2}$ and the decoherence model, given by Eq. (28), was fitted simultaneously to all bins (the same $W, \mathcal{V}_0$, different $\tau$). Resulting $\tau(|\mathbf{k}|)$ corresponds to the solid line (Data) and the shaded area depicting a standard deviation. Dashed line (Fit) represents the fit of $\tau(|\mathbf{k}|) = \gamma/|\mathbf{k}|$ with obtained $\gamma = 5.98 \times 10^3 \ \mu\text{s rad mm}^{-1}$ corresponding to a temperature of 52 μK. Other fit parameters are consistent with previous results $W = 0.15 \pm 0.01$, $\mathcal{V}_0 = 0.94 \pm 0.01$.

We have demonstrated CHSH inequality violation by more than 3 standard deviations with the Bell parameter reaching $\mathcal{S} = 2.60 \pm 0.19$. Via wavevector-resolved Bell state measurement (BSM) with varying storage times, we established experimentally and with a theoretical model the BSM visibility—quantity proportional to the Bell parameter under the optimal choice of bases—behavior with wavevector and temporal resolution. After 45 μs storage time, ca. 50% of modes still indicate CHSH violation, while after 60 μs it amounts to around 10%. Importantly, with single-copy distillation protocols[65], even slightly entangled states can be probabilistically distilled into Bell states, hence the BSM visibility averaged over the modes—which reflects the average ebit content of a generated state—is a significant figure of merit. In our experiment, it has fallen merely to around 50% at 60 μs storage time, as compared to the initial ca. 80% for immediate readout. With the current experimental parameters, we estimate the probability to generate and detect at least one atom-photon Bell state across all pairs of modes to be $1 - (1 - \eta\chi)^M \approx 35\%$ (89% when only the detection and not the filtering system efficiency is contained in $\eta$).

Quantum repeater networks constitute the most versatile application of entanglement amending the fundamental exponential loss of photonic channels. While the performance and feasibility of WV-MUX-QM platform for quantum repeater networks has been discussed elsewhere[51], we note here that obtaining hybrid entanglement is essential for two-photon protocols[58–60] which solve the phase stability issues inherent to the DLCZ protocol[25]. From the perspective of feasible quantum repeaters, the 800 nm photons are sub-optimal having an order of magnitude higher fiber-transmission losses than telecom; however, conversion to telecom photons as well as light-atom interfaces at telecom wavelengths have been demonstrated[2,66–68]. Furthermore, quantum repeater protocols generally do not require phase stability between different modes, rendering free-space transmission a viable alternative.

Importantly, WV-MUX-QM platform offers versatility beyond entanglement generation, such as intramemory processing of stored spin-wave states[69]—including operations on non-classical single-excitation states[50], continuous variable and temporal processing[70,71] or ultra-narrow band temporal imaging[72]. Furthermore, the rich atomic structure of alkali atoms may be employed to design various light-matter interfaces suited for a particular application[73].

Finally, we note that our experimental demonstration is far from the ultimate capabilities of wavevector-multiplexed memories. Cold atomic memories with higher coherence times[37] and read-out efficiencies[31] has been demonstrated. The wavevector multiplexing principle could be applied with atomic clouds captured in a dipole trap or an optical lattice, which can prolong the memory lifetime to the sub-second regime[36,62]. However, the simplest improvement would be to observe a larger angular range of photonic emission from the memory. As estimated in our theoretical work[51], the number of modes attainable with standard optical components and commercial CMOS image sensors would reach over 5000 enabling feasible high-rate entanglement generation.

## Methods

**Wavevector-dependent phase**. The ability to shape the wave-vector dependent phase $\varphi_w(\mathbf{k}_w)$, $\varphi_r(\mathbf{k}_r)$ opens vast possibilities. For instance, consider the most-strongly correlated write-out and read-out wavevectors $\mathbf{k}_w = \mathbf{k}_r = \mathbf{k}$. By selecting $\varphi(\mathbf{k}) \equiv \varphi_w(\mathbf{k}) + \varphi_r(\mathbf{k})$ periodically equal to either $\pi$ or 0 on a rectangular grid, we generate either a $\Phi_-$ or $\Phi_+$ Bell-state, respectively, depending on the wavevector $\mathbf{k}$. Interestingly, the wavevector of the scattered write-out photon $\mathbf{k}$ is inherently random and selected by the process of spontaneous Raman scattering. Such a quantum-random sampling of generated states is equivalent to a quantum-random choice of measurement bases for Alice and Bob in the Bell setting (since either the states or the bases may be equivalently altered). After the measurement, the central WV-MUX-QM can reveal the exact phase profile $\varphi(\mathbf{k})$ that was employed, allowing Alice and Bob to interpret their results.

**The WV-MUX-QM platform**. The WV-MUX-QM is based on an elongated (0.5 mm × 0.5 mm × 7 mm), high-optical-depth ensemble of rubidium-87 atoms prepared in magnetooptical trap (MOT). The optical depth (OD) amounts to 150 at the $F=1, m_F=1 \rightarrow F=2, m_F=2$ transition of D2 line, which corresponds to OD = 25 at the Write laser transition (in resonance). To define the quantization axis, the cloud is kept in a constant (1 Gauss) magnetic field oriented along the propagation axis (z-axis). The MOT loading time is set to 2 ms which includes compression stage (700 µs) and polarization gradient cooling, with magnetic gradient switched off (300 µs). After MOT loading stage the atoms are optically pumped to the $F=1, m_F=-1$ state using 70 µs-long hyperfine pump pulse (resonant with the $F=2 \rightarrow F=2$ transition of D2 line), that illuminates the cloud from four sides (along cooling beams) and has 15 mW power in total. The Zeeman sublevel pumping is achieved by illuminating the atoms along the z-axis by a circularly polarized laser beam (resonant with $F=1 \rightarrow F=1$ transition of D2 line) for 55 µs from the beginning of the hyperfine pump pulse. We estimate the efficiency of the Zeeman pumping to be about 70%[43]. Longer hyperfine pump pulse duration combined with additional (5 mW, 75 µs) "clearing" pulse of the Read laser guarantees, that the storage level ($F=2, m_F=1$) is emptied before the quantum memory protocol. To generate the two-mode squeezed state of spin-wave and a write-out photon, we use Raman interaction in the Λ scheme. The write-out photon and spin-wave pairs are generated using a 30 MHz red-detuned $\sigma_+$-polarized Write laser pulse (300 ns) on $F=1 \rightarrow F'=2$ transition of D2 line. The Write laser power is chosen to provide desired pair generation probability $\chi \approx 0.01$, which in our case corresponds to about 10 µW. Since we filter the write-out photons to have orthogonal polarization to the laser photons, the spin waves of our interest are created between the $F=1, m_F=-1$ and $F=2, m_F=1$ states. To convert (read-out) the spin waves to read-out photons we use $\sigma_-$-polarized Read laser pulse (300 ns) tuned to the $F=2 \rightarrow F'=2$ transition of the D1 line. The Read laser power is chosen to give the readout pulse duration of approx. 200 ns, which corresponds to about 100 µW. The double-Λ configuration (consisting of D1 and D2 lines of rubidium-87) allows us to efficiently filter write-out and read-out photons from back-emissions, that may lead to uncorrelated detection events and spoil the measurements. Finally, to provide the best signal-to-noise ratio, the write-out and read-out photons pass through narrowband atomic filters. The filters are glass cells containing optically pumped $^{87}$Rb that absorb residual laser light (which is separated from photons by 6.8 GHz), while being transparent to the signal (write-out or read-out) photons (see Parniak et al. for details of the filtering system[43]).

**Single-photon sensitive I-sCMOS camera**. Generation of Bell states across many modes requires efficient detection of single-photons with a high spatial resolution. High number of modes—several hundreds—render arrays of single-mode single-photon detectors practically unfeasible; however, recent development in single-photon cameras shows the detection technology is sufficient for near-term applications[63,74]. Among single-photon camera solution, the most commonly employed are intensified CCD (ICCD) and electron-multiplying CCD (EMCCD). Neither of those solutions achieves acquisition times on the order of a few µs, which would allow real-time response to detected photons required in many protocols such as quantum repeater operation. Additionally, EMCCD cameras have a high read-out noise resulting in high dark-count rates and consequently deteriorating the fidelity of the generated entanglement. Recently commercialized, intensified scientific CMOS (I-sCMOS) cameras solve those issues to a great extent. Additionally, so-called quanta image sensors[74] offer further improvements, especially in terms of quantum efficiency.

In our experiment, we employ a custom I-sCMOS camera which involves a 2-stage image intensifier (luminous gain of $5 \times 10^6$) imaged by a relay lens (f-number of 1.1, magnification of −0.44) onto a 5.5 Mpx (mega pixel), 2560 × 2160 px CMOS sensor (effective pixel pitch ca. 15 µm accounting for relay lens magnification). Image intensifier involves a GaAs photocathode which around 800 nm offers the overall detection efficiency of ca. 20%. Active multiplexing, required in many applications[51] would inherently need real-time feedback from the camera, which has not been implemented in the current experiment; however, it would be possible with one of the modern fast CMOS sensors.

*Mode detection cross-talk*. With the high resolution of single-photon cameras, the probability for two or more generated write-out photons to share the same (up to the camera resolution) central wavevector is given by $\approx 1 - n!/(n - \lceil \chi M \rceil)! \times 1/n^{\lceil \chi M \rceil}$, where $n = 130 \times 160$ is the number of observed pixels (px). In our case ($\chi \approx 0.01$) the probability is $4.8 \times 10^{-4}$ and we will neglect such cases. Similarly, if we focus on a single jth mode with a detected write-out photon, the probability of registering a read-out photon coming from another mode in the 3σ radius around the most-likely position of jth read-out can be approximated as $1 - [1 - \chi \eta_r \eta \times \pi(3\sigma)^2/n]^{M-1}$, where σ is the mode size in pixels, $\eta_r \approx 40\%$ corresponds to the read-out efficiency and $\eta \approx 8\%$ to the filtering and detection system efficiency. In our case the probability amounts to around 0.17%. Therefore, in most cases, we can consider a single isolated jth mode with the results being valid for any j.

The used formulae have been derived in the Supplementary Material Sec. S6.

**Visibility model**. Assuming low average number of detected photons per experiment $\bar{n} \ll 1$, we can approximate the photon number cross-correlation function as:

$$g^{(2)}(\mathbf{k}_r, \mathbf{k}_w) \approx p_{w,r}/(p_w p_r), \quad (32)$$

where $p_{w,r} \equiv p_{w,r}(\mathbf{k}_r, \mathbf{k}_w)$ is the single-experiment probability of observing a coincidence between a write-out and read-out photon and $p_w \equiv p_w(\mathbf{k}_w)$ ($p_r \equiv p_r(\mathbf{k}_r)$) denotes the marginal probability of observing a write-out (read-out) photon. The coincidence probability can be written as $p_{w,r} = g^{(2)} p_w p_r$, with $g^{(2)} = g^{(2)}(\mathbf{k}_r, \mathbf{k}_w)$. For a given point $(\mathbf{k}_r, \mathbf{k}_w)$ in the wavevector space, we measure the BSM visibility by comparing the number of coincidences during measurement set up for maximally constructive (+) or destructive (−) interference. The numbers of coincidences serve as probability estimates and hence give the visibility as

$$\mathcal{V}(\mathbf{k}_r, \mathbf{k}_w) = (p_{w,r}^{(+)} - p_{w,r}^{(-)})/(p_{w,r}^{(+)} + p_{w,r}^{(-)}). \quad (33)$$

In (−) settings only noise coincidences are registered i.e. $p_{w,r}^{(-)} = p_w p_r$ while $p_{w,r}^{(+)} = p_{w,r} = g^{(2)} p_w p_r$. This directly gives us Eq. (20).

*Storage-time dependence*. Let us first consider the visibility model as given by Eq. (21) and under experimentally verified assumptions of negligible write-out noise $\tilde{B}_w \ll \chi$:

$$\mathcal{V} \approx \left[1 + \frac{2\chi}{F(n)} + \frac{2\tilde{B}_r/\eta_r}{F(n)}\right]^{-1}. \quad (34)$$

We shall further employ a model for the read-out noise:

$$\tilde{B}_r(t) = \tilde{B}_r(0) + \tilde{B}_r^{(\chi)} \chi + [\tilde{B}_r(\infty) - \tilde{B}_r(0)] \times [1 - \exp(-t/\tau_B)], \quad (35)$$

which has been verified in an additional measurement, as described in the Supplementary Material Sec. S4. Furthermore, with the calibrated values of $\tau_B \approx 13$ µs and $\tilde{B}_r(\infty) \approx 5 \times \tilde{B}_r(0)$ we can approximate

$$\tilde{B}_r(t) \approx \tilde{B}_r(\infty) + \tilde{B}_r^{(\chi)} \chi \equiv B \quad (36)$$

with little error. Denoting

$$\mathcal{V}_0 = 1/(1 + 2\chi/F(n)), \quad (37)$$

$$W = 2F(n)B/[\eta_r(0) \times (2\chi + F(n))^2], \quad (38)$$

and using the decohrence model of Eq. (26) we get the Eq. (28).

*Different wavevectors for superimposed H, V modes.* Here we quantify the effects of different decoherence rates on the generated states. Let us denote

$$|\mathbf{k_H}|^2 = y^2 + x^2, \quad (39)$$

and

$$|\mathbf{k_V}|^2 = (y_{max} - y)^2 + x^2 = |\mathbf{k_H}|^2 + y_{max}^2 - 2y_{max}y, \quad (40)$$

where $y_{max}$ corresponds to the maximal $y$ wavevector component entering the MZI. We shall focus on a single mode and modify the state given by Eq. (9) so that $H$ and $V$ parts have different (modulus) coefficients:

$$|\Phi(\varphi,t)\rangle = \frac{[c_H(t)|H\rangle_r|H\rangle_w + c_V(t)\exp(i\varphi)|V\rangle_r|V\rangle_w]}{\sqrt{c_H^2(t) + c_V^2(t)}}, \quad (41)$$

with

$$c_P^2(t) = \exp(-t^2|\mathbf{k_P}|^2/\gamma^2); \; P \in \{H,V\}. \quad (42)$$

Assuming $|\mathbf{k_H}| > |\mathbf{k_V}|$ the temporal evolution of the state in Eq. (41) brings it closer to $|H\rangle_r|H\rangle_w$. If we consider a reduced state e.g., of write-out only, the evolution moves the state from the equator of the Bloch sphere to the pole. Intuitively, as the polarization measurement projects the state on some axis in the equator plane, the visibility will be reduced. Direct calculation yields

$$\mathrm{Tr}\Big[(\Pi_{\xi_w} \otimes \Pi_{\xi_r})|\Phi(\varphi,t)\rangle\langle\Phi(\varphi,t)|\Big]$$
$$= \tilde{\mathcal{V}}(t)\cos(\xi_r + \xi_w + \varphi) \quad (43)$$

with

$$\tilde{\mathcal{V}}(t) = 1/\cosh[t^2/(2\gamma^2) \times (y_{max}^2 - 2y_{max}y)]. \quad (44)$$

Importantly, no assumption on the actual form of $\tau(\mathbf{k})$ need to be made to give an expression for $\tilde{\mathcal{V}}(t)$, as long as Eq. (26) holds. Let us denote $k_{min} = \min(|\mathbf{k_H}|, |\mathbf{k_V}|)$, $k_{max} = \max(|\mathbf{k_H}|, |\mathbf{k_V}|)$. Since longer wavevectors are associated with faster decoherence, let us also denote $\tau_{min} = \tau(k_{max})$, $\tau_{max} = \tau(k_{min})$ and $\Delta = \tau_{min}^{-1} - \tau_{max}^{-1}$. This way, we arrive at Eq. (27).

Interestingly, MZI configured to give different $c_H(t)$ and $c_V(t)$ could be used to demonstrate a violation of the so-called tilted Bell inequalities[75] by generating a family of states with a varying degree of entanglement, across the memory modes. The only required modification in the experimental setup would be to remove quarter waveplates, depicted in Fig. 1, which would amount to performing BSM with polarization operators of the form $\hat{\sigma}_z \cos\xi + \hat{\sigma}_x \sin\xi$ corresponding to the projection on the Bloch sphere's meridian.

## Data availability
Data presented in Figs. 2–5 have been deposited in RepOD Repository for Open Data at https://doi.org/10.18150/RJBMOD. Any other data that support the findings of this study are available from the corresponding author upon reasonable request.

## Code availability
Data was analyzed using custom photon analysis code available at https://github.com/Michuu/photonpacket. Other custom code used in data analysis is available from the corresponding author upon reasonable request.

## References

1. Abruzzo, S., Kampermann, H. & Bruß, D. Measurement-device-independent quantum key distribution with quantum memories. *Phys. Rev. A* **89**, 012301 (2014).
2. Chang, W. et al. Long-distance entanglement between a multiplexed quantum memory and a telecom photon. *Phys. Rev. X* **9**, 041033 (2019).
3. Dam, S. B. V., Humphreys, P. C., Rozpędek, F., Wehner, S. & Hanson, R. Multiplexed entanglement generation over quantum networks using multi-qubit nodes. *Quantum Sci. Technol.* **2**, 034002 (2017).
4. Kimble, H. J. The quantum internet. *Nature* **453**, 1023–1030 (2008).
5. Krovi, H. et al. Practical quantum repeaters with parametric down-conversion sources. *Appl. Phys. B* **122**, 52 (2016).
6. Lee, J. & Kim, M. S. Entanglement Teleportation via Werner States. *Phys. Rev. Lett.* **84**, 4236–4239 (2000).
7. Li, C. et al. Quantum communication between multiplexed atomic quantum memories. *Phys. Rev. Lett.* **124**, 240504 (2020).
8. Munro, W. J., Harrison, K. A., Stephens, A. M., Devitt, S. J. & Nemoto, K. From quantum multiplexing to high-performance quantum networking. *Nat. Photonics* **4**, 792–796 (2010).
9. Muralidharan, S. et al. Optimal architectures for long distance quantum communication. *Sci. Rep.* **6**, 20463 (2016).
10. Panayi, C., Razavi, M., Ma, X. & Lütkenhaus, N. Memory-assisted measurement-device-independent quantum key distribution. *New J. Phys.* **16**, 043005 (2014).
11. Razavi, M., Thompson, K., Farmanbar, H., Piani, M. & Lütkenhaus, N. in *Quantum Communications Realized II* vol. 7236 (eds. Arakawa, Y. et al.) (SPIE, 2009).
12. Simon, C. Towards a global quantum network. *Nat. Photonics* **11**, 678–680 (2017).
13. Wehner, S., Elkouss, D. & Hanson, R. Quantum internet: a vision for the road ahead. *Science* **362**, eaam9288 (2018).
14. Vinay, S. E. & Kok, P. Practical repeaters for ultralong-distance quantum communication. *Phys. Rev. A* **95**, 52336 (2017).
15. Yang, W. J. & Wang, X. B. Heralded quantum entanglement between distant matter qubits. *Sci. Rep.* **5**, 10110 (2015).
16. Ekert, A. K. in *Quantum Measurements in Optics* (eds. Tombesi, P. & Walls, D. F.) 413–418 (Springer, 1992).
17. Pirandola, S. et al. High-rate measurement-device-independent quantum cryptography. *Nat. Photonics* **9**, 397–402 (2015).
18. Pirandola, S. et al. Advances in quantum cryptography. *Adv. Opt. Photonics* **12**, 1012–1236 (2020).
19. Zhang, W. et al. Quantum secure direct communication with quantum memory. *Phys. Rev. Lett.* **118**, 220501 (2017).
20. Zwerger, M., Pirker, A., Dunjko, V., Briegel, H. J. & Dür, W. Long-range big quantum-data transmission. *Phys. Rev. Lett.* **120**, 030503 (2018).
21. Lim, Y. L., Beige, A. & Kwek, L. C. Repeat-until-success linear optics distributed quantum computing. *Phys. Rev. Lett.* **95**, 030505 (2005).
22. Bunandar, D. et al. Metropolitan quantum key distribution with silicon photonics. *Phys. Rev. X* **8**, 021009 (2018).
23. Da Lio, B. et al. Experimental demonstration of the DPTS QKD protocol over a 170 km fiber link. *Appl. Phys. Lett.* **114**, 011101 (2020).
24. Briegel, H.-J., Dür, W., Cirac, J. I. & Zoller, P. Quantum repeaters: the role of imperfect local operations in quantum communication. *Phys. Rev. Lett.* **81**, 5932–5935 (1998).
25. Duan, L. M., Lukin, M. D., Cirac, J. I. & Zoller, P. Long-distance quantum communication with atomic ensembles and linear optics. *Nature* **414**, 413–8 (2001).
26. Michael, M. H. et al. New class of quantum error-correcting codes for a bosonic mode. *Phys. Rev. X* **6**, 031006 (2016).
27. Li, Z.-D. et al. Experimental quantum repeater without quantum memory. *Nat. Photonics* **13**, 644–648 (2019).
28. Jiang, N. et al. Experimental realization of 105-qubit random access quantum memory. *npj Quantum Inf.* **5**, 28 (2019).
29. Yuan, Z. S. et al. Experimental demonstration of a BDCZ quantum repeater node. *Nature* **454**, 1098–1101 (2008).
30. Bao, X.-H. et al. Efficient and long-lived quantum memory with cold atoms inside a ring cavity. *Nat. Phys.* **8**, 517–521 (2012).
31. Cho, Y.-W. et al. Highly efficient optical quantum memory with long coherence time in cold atoms. *Optica* **3**, 100–107 (2016).
32. Sabooni, M., Li, Q., Kröll, S. & Rippe, L. Efficient quantum memory using a weakly absorbing sample. *Phys. Rev. Lett.* **110**, 133604 (2013).
33. Wang, Y. et al. Efficient quantum memory for single-photon polarization qubits. *Nat. Photonics* **13**, 346–351 (2019).
34. Cao, M., Hoffet, F., Qiu, S., Sheremet, A. S. & Laurat, J. Efficient reversible entanglement transfer between light and quantum memories. *Optica* **7**, 1440 (2020).
35. Nunn, J. et al. Quantum memory in an optical lattice. *Phys. Rev. A* **82**, 022327 (2010).
36. Yang, S. J., Wang, X. J., Bao, X. H. & Pan, J. W. An efficient quantum light-matter interface with sub-second lifetime. *Nat. Photonics* **10**, 381–384 (2016).
37. Zhao, R. et al. Long-lived quantum memory. *Nat. Phys.* **5**, 100–104 (2009).
38. Zhao, B. et al. A millisecond quantum memory for scalable quantum networks. *Nat. Phys.* **5**, 95–99 (2009).
39. Collins, O. A., Jenkins, S. D., Kuzmich, A. & Kennedy, T. A. B. Multiplexed memory-insensitive quantum repeaters. *Phys. Rev. Lett.* **98**, 060502 (2007).
40. Simon, C., De Riedmatten, H. & Afzelius, M. Temporally multiplexed quantum repeaters with atomic gases. *Phys. Rev. A* **82**, 010304 (2010).
41. Tian, L. et al. Spatial multiplexing of atom-photon entanglement sources using feedforward control and switching networks. *Phys. Rev. Lett.* **119**, 130505 (2017).
42. Yang, T.-S. et al. Multiplexed storage and real-time manipulation based on a multiple degree-of-freedom quantum memory. *Nat. Commun.* **9**, 3407 (2018).
43. Parniak, M. et al. Wavevector multiplexed atomic quantum memory via spatially-resolved single-photon detection. *Nat. Commun.* **8**, 2140 (2017).
44. Pu, Y. F. et al. Experimental realization of a multiplexed quantum memory with 225 individually accessible memory cells. *Nat. Commun.* **8**, 15359 (2017).

45. Vernaz-Gris, P., Huang, K., Cao, M., Sheremet, A. S. & Laurat, J. Highly-efficient quantum memory for polarization qubits in a spatially-multiplexed cold atomic ensemble. *Nat. Commun.* **9**, 363 (2018).
46. Tang, J. S. et al. Storage of multiple single-photon pulses emitted from a quantum dot in a solid-state quantum memory. *Nat. Commun.* **6**, 8652 (2015).
47. Gündoğan, M., Ledingham, P. M., Kutluer, K., Mazzera, M. & De Riedmatten, H. Solid state spin-wave quantum memory for time-bin qubits. *Phys. Rev. Lett.* **114**, 230501 (2015).
48. Wen, Y. et al. Multiplexed spin-wave-photon entanglement source using temporal multimode memories and feedforward-controlled readout. *Phys. Rev. A* **100**, 012342 (2019).
49. Sinclair, N. et al. Spectral multiplexing for scalable quantum photonics using an atomic frequency comb quantum memory and feed-forward control. *Phys. Rev. Lett.* **113**, 053603 (2014).
50. Parniak, M. et al. Quantum optics of spin waves through ac Stark modulation. *Phys. Rev. Lett.* **122**, 063604 (2019).
51. Lipka, M., Mazelanik, M. & Parniak, M. Entanglement distribution with wavevector-multiplexed quantum memory. https://arxiv.org/abs/2007.00538 (2020).
52. Richardson, D. J., Fini, J. M. & Nelson, L. E. Space-division multiplexing in optical fibres. *Nat. Photonics* **7**, 354–362 (2013).
53. Ding, Y. et al. High-dimensional quantum key distribution based on multicore fiber using silicon photonic integrated circuits. *npj Quantum Inf.* **3**, 25 (2017).
54. Liao, S. K. et al. Long-distance free-space quantum key distribution in daylight towards inter-satellite communication. *Nat. Photonics* **11**, 509–513 (2017).
55. de Riedmatten, H. et al. Direct measurement of decoherence for entanglement between a photon and stored atomic excitation. *Phys. Rev. Lett.* **97**, 113603 (2006).
56. Matsukevich, D. N. et al. Entanglement of a photon and a collective atomic excitation. *Phys. Rev. Lett.* **95**, 040405 (2005).
57. Wang, S. et al. Long-lived and multiplexed atom-photon entanglement interface with feed-forward-controlled readouts, https://arxiv.org/abs/2006.05631 (2020).
58. Chen, Z.-B., Zhao, B., Chen, Y.-A., Schmiedmayer, J. & Pan, J.-W. Fault-tolerant quantum repeater with atomic ensembles and linear optics. *Phys. Rev. A* **76**, 022329 (2007).
59. Zhao, B., Chen, Z.-B., Chen, Y.-A., Schmiedmayer, J. & Pan, J.-W. Robust creation of entanglement between remote memory qubits. *Phys. Rev. Lett.* **98**, 240502 (2007).
60. Jiang, L., Taylor, J. M. & Lukin, M. D. Fast and robust approach to long-distance quantum communication with atomic ensembles. *Phys. Rev. A* **76**, 012301 (2007).
61. Dąbrowski, M. et al. Certification of high-dimensional entanglement and Einstein-Podolsky-Rosen steering with cold atomic quantum memory. *Phys. Rev. A* **98**, 42126 (2018).
62. Dudin, Y. O., Li, L. & Kuzmich, A. Light storage on the time scale of a minute. *Phys. Rev. A* **87**, 031801 (2013).
63. Lipka, M., Parniak, M. & Wasilewski, W. Microchannel plate cross-talk mitigation for spatial autocorrelation measurements. *Appl. Phys. Lett.* **112**, 211105 (2018).
64. Wootters, W. K. Entanglement of formation of an arbitrary state of two qubits. *Phys. Rev. Lett.* **80**, 2245–2248 (1998).
65. Wang, Z.-W. et al. Experimental entanglement distillation of two-qubit mixed states under local operations. *Phys. Rev. Lett.* **96**, 220505 (2006).
66. Radnaev, A. G. et al. A quantum memory with telecom-wavelength conversion. *Nat. Phys.* **6**, 894–899 (2010).
67. Chanelière, T. et al. Quantum telecommunication based on atomic cascade transitions. *Phys. Rev. Lett.* **96**, 093604 (2006).
68. Albrecht, B., Farrera, P., Fernandez-Gonzalvo, X., Cristiani, M. & de Riedmatten, H. A waveguide frequency converter connecting rubidium-based quantum memories to the telecom C-band. *Nat. Commun.* **5**, 3376 (2014).
69. Leszczyński, A. et al. Spatially resolved control of fictitious magnetic fields in a cold atomic ensemble. *Opt. Lett.* **43**, 1147 (2018).
70. Mazelanik, M., Parniak, M., Leszczyński, A., Lipka, M. & Wasilewski, W. Coherent spin-wave processor of stored optical pulses. *npj Quantum Inf.* **5**, 22 (2019).
71. Lipka, M., Leszczyński, A., Mazelanik, M., Parniak, M. & Wasilewski, W. Spatial spin-wave modulator for quantum-memory-assisted adaptive measurements. *Phys. Rev. Appl.* **11**, 034049 (2019).
72. Mazelanik, M., Leszczyński, A., Lipka, M., Parniak, M. & Wasilewski, W. Temporal imaging for ultra-narrowband few-photon states of light. *Optica* **7**, 203–208 (2020).
73. Mazelanik, M., Leszczyński, A., Lipka, M., Wasilewski, W. & Parniak, M. Superradiant parametric conversion of spin waves. *Phys. Rev. A* **100**, 053850 (2019).
74. Ma, J., Masoodian, S., Starkey, D. A. & Fossum, E. R. Photon-number-resolving megapixel image sensor at room temperature without avalanche gain. *Optica* **4**, 1474 (2017).
75. Christensen, B. G., Liang, Y.-C., Brunner, N., Gisin, N. & Kwiat, P. G. Exploring the limits of quantum Nonlocality with entangled photons. *Phys. Rev. X* **5**, 041052 (2015).

### Acknowledgements
This work has been funded by the Ministry of Education and Science (Poland) grants no. DI2016 014846, DI2018 010848, by the National Science Centre (Poland) grants no. 2016/21/B/ST2/02559, 2017/25/N/ST2/01163, 2017/25/N/ST2/00713, by the Foundation for Polish Science MAB/2018/4 "Quantum Optical Technologies" project and by the Office of Naval Research (USA) grant no. N62909-19-1-2127. The "Quantum Optical Technologies" project is carried out within the International Research Agendas programme of the Foundation for Polish Science co-financed by the European Union under the European Regional Development Fund. M.P. has been also supported by the Foundation for Polish Science via the START scholarship. We thank K. Banaszek for the generous support and J. Kołodyński for insightful discussions.

### Author contributions
M.M. and A.L. performed the measurements assisted by M.L.; M.L. analysed the data assisted by M.M.; M.L. performed the theoretical analysis and wrote the manuscript assisted by M.M. and M.P.; M.L., M.M., A.L., M.P., and W.W. contributed to building and calibration of the experimental setup; M.L., M.M., and M.P. conceived the scheme and designed the experiment; W.W. and M.P. supervised the project.

### Competing interests
The University of Warsaw has filed a related Polish Patent Application P.434142 entitled "System for generating entangled photon pairs of multimode quantum memory for regeneration of a quantum signal at a long distance, a method for generating entangled photon pairs of multimode quantum memory for regeneration of a quantum signal at a long distance" invented by W. Wasilewski, M. Lipka, M. Mazelanik, M. Parniak, K. Zdanowski, A. Ostasiuk, and A. Leszczyński. The unpublished patent application is pending. The authors declare no other competing interests.

### Additional information
**Supplementary information** The online version contains supplementary material available at https://doi.org/10.1038/s42005-021-00551-1.

**Correspondence** and requests for materials should be addressed to M.P.

**Reprints and permission information** is available at http://www.nature.com/reprints

**Publisher's note** Springer Nature remains neutral with regard to jurisdictional claims in published maps and institutional affiliations.